\documentclass[12pt]{article}
\usepackage[utf8]{inputenc}
\usepackage[english]{babel}
\usepackage[centertags]{amsmath}
\usepackage{amssymb}
\usepackage{bm}
\usepackage{amsbsy}
\usepackage{graphicx}
\usepackage{appendix}
\usepackage{mathrsfs}
\usepackage{epsfig}
\usepackage{color}
\usepackage{euscript}
\usepackage{ulem}
\usepackage{multirow}
\usepackage{cite}
\usepackage{textcomp}

\definecolor{darkblue}{rgb}{0,0,1}

\definecolor{darkraspberry}{rgb}{0.53,0.15,0.34}

\definecolor{dgreen}{rgb}{0,0.6,0}

\definecolor{darkred}{rgb}{0.8,0,1}

\definecolor{purple}{rgb}{0.5,0.0,0.5}

\usepackage{jheppubm}

\newcommand{\nn}{\nonumber}
\newcommand{\be}{\begin{equation}}
\newcommand{\ee}{\end{equation}}
\newcommand{\bea}{\begin{eqnarray}}
\newcommand{\eea}{\end{eqnarray}}

\newcommand{\fb}{\mathfrak{b}}

\newcommand{\fg}{\mathfrak{g}}

\newcommand{\cA}{\cal A}
\newcommand{\cL}{\cal L}
\newcommand{\cS}{\cal S}

\numberwithin{equation}{section}

\title{Phase Diagram Magnetic Features \\ of Holographic Anisotropic Model \\ for $z^4$-term Heavy Quarks}

\author{
  Kristina Rannu$^{a}$
  }

\affiliation{
$^a$Peoples Friendship University of Russia,\\ Miklukho-Maklaya
str. 6, 117198, Moscow, Russia
}

\emailAdd{rannu-ka@rudn.ru}

\abstract{Thermodynamical features depending on magnetic field of the previously found five-dimensional anisotropic holographic solution for heavy quarks supported by Einstein-dilaton-three-Maxwell action with extended warp factor and improved coupling function for the Maxwell field providing chemical potential are studied. Direct/inverse magnetic catalysis behavior is considered. Phase diagram, composed of the 1-st order phase transition and temporal Wilson loops, corresponding to the stable direct magnetic catalysis scenario is presented. String tension behavior in magnetic field is discussed.}


\begin{document}

\maketitle

\section{Introduction}

This work purpose is to extent the confinement-deconfinement phase transition investigation in the previously constructed holographic model of the hot dense anisotropic quark-gluon plasma (QGP) in magnetic field, produced in heavy-ion collisions (HIC) \cite{Arefeva:2023jjh}. The main objective of the model was to get the effect of direct magnetic catalysis, i.e. the increase of the phase transition temperature in stronger magnetic field \cite{Miransky:2002rp, Shovkovy:2012zn}, in holographic description of heavy quarks confinement/deconfinement scenario with external magnetic field \cite{AR-2018, ARS-2019, ARS-Heavy-2020}. The model is well developed already by the running coupling and $\beta$-function considerations \cite{Arefeva:2024vom, Arefeva:2024xmg, Arefeva:2024poq, Slepov:2024vjt, Usova:2024qmf, Hajilou:2024swf, Arefeva:2025xtz, Arefeva:2025okg}.

In this work temporal Wilson loops are considered, thus completing the phase diagram as the interplay of the 1-st order phase transition and the so-called crossover. The obtained picture is complemented by the string tension behavior and investigated in the presence of the magnetic field.

The paper is organised as follows. In Sect.~\ref{model} the holographic model for hot dense anisotropic QGP of heavy quarks is described. Is Sect.~\ref{thermodynamics} the thermodynamical properties of the model and the 1-st order phase transition depending on the magnetic field are discussed. In Sect.~\ref{twl} temporal Wilson loops (TWL) are calculated, the crossover and the string tension originating from the TWL behavior are presented. In Sect.~\ref{conclusions} the obtained phase diagram is investigated to draw conclusions about its magnetic properties. Details of the solution and some additional plots are presented in Appendix~\ref{appendixA}.

\section{Model}\label{model}

The model considered was obtained as a solution for the Lagrangian in the Einstein frame \cite{Arefeva:2023jjh}:
\begin{gather}
  {\cL} = \sqrt{-g} \left[ R 
    - \cfrac{f_0(\phi)}{4} \, F_0^2 
    - \cfrac{f_1(\phi)}{4} \, F_1^2
    - \cfrac{f_3(\phi)}{4} \, F_3^2
    - \cfrac{1}{2} \, \partial_{\mu} \phi \, \partial^{\mu} \phi
    - V(\phi) \right], \label{eq:2.01}  \\
  \phi = \phi(z), \label{eq:2.02} \\
  \begin{split}
    \mbox{Electric anzats $F_0$:} \quad
    &A_0 = A_t(z), \quad A_{i = 1,2,3,4} = 0, \\
    \mbox{Magnetic anzats $F_k$:} \quad
    &F_1 = q_1 \, dx^2 \wedge dx^3, \quad 
    F_3 = q_3 \, dx^1 \wedge dx^2.
  \end{split}\label{eq:2.03}
\end{gather}
where $\phi = \phi(z)$ is the scalar field, $f_0(\phi(z)) = f_0(z)$, $f_1(\phi(z)) = f_1(z)$ and $f_3(\phi(z)) =
f_3(z)$ are the coupling functions associated with the Maxwell
fields $A_{\mu}$, $F_1$ and $F_3$ correspondingly, $q_1$ and $q_3$ are
constants ``charges'' and $V(\phi(z)) = V(z)$ is the scalar field potential. 

The solution is given by the following ansatz:
\begin{gather}
  ds^2 = \cfrac{L^2}{z^2} \, \fb(z) \left[
    - \, g(z) \, dt^2 + dx_1^2 
    + \left( \cfrac{z}{L} \right)^{2-\frac{2}{\nu}} dx_2^2
    + e^{c_B z^2} \left( \cfrac{z}{L} \right)^{2-\frac{2}{\nu}} dx_3^2
    + \cfrac{dz^2}{g(z)} \right] \! , \label{eq:2.04} \\
  \fb(z) = e^{2{\cA}(z)}, \nn
\end{gather}
where $L$ is the AdS-radius, $\fb(z)$ is the warp factor set by
${\cA}(z)$, $g(z)$ is the blackening function, $\nu$ is the parameter
of primary anisotropy, caused by non-symmetry of heavy-ion collision
(HIC), and $c_B$ is the coefficient of secondary anisotropy related to
the magnetic field $F^{(3)}$. Choice of ${\cA}(z)$
determines the heavy/light quarks description of the model. In
previous works we considered ${\cA}(z) = - \, c z^2/4$ for heavy
quarks \cite{AR-2018, ARS-2019, ARS-Heavy-2020} and ${\cA}(z) = - \, a
\, \ln (b z^2 + 1)$ for light-quarks \cite{ARS-Light-2020, ARS-Light-2022, Arefeva:2022bhx}.

To provide the the direct magnetic catasysis effect the warp factor and the ``electric'' Maxwell field coupling function are chosen as \cite{Arefeva:2023jjh} 
\begin{gather}
  \fb(z) = e^{2{\cA}(z)} = e^{ - \frac{c z^2}{2} - 2 (p-c_B \, q_3)
    z^4}, \label{eq:4.42} \\
  f_0 = e^{-(R_{gg}+\frac{c_B q_3}{2})z^2} \,
    \cfrac{z^{-2+\frac{2}{\nu}}}{\sqrt{\fb}}, \label{eq:4.27}
\end{gather}
where $c = 4 R_{gg}/3$, $R_{gg} = 1.16$ and $p = 0.273$ to fit the Regge spectra of meson \cite{He:2020fdi}. The solution (\ref{eq:4.28}--\ref{eq:4.48}) is given in Appendix~\ref{appendixA}.

There are two magnetic field parameters in this model. The metric coefficient $c_B$ characterizes the secondary anisotropy, caused by magnetic field and related to the non-centrality of HIC. The magnetic field magnitude $q_3$ is contained in the warp factor and enters the blackening function, thus having its influence on the solution. In general case these parameters are independent from each other, and the model magnetic properties strongly depend on their interplay.

The most obvious particular cases of magnetic field parametrization are fixed $q_3$ (A), fixed $c_B$ (B), $q_3 = \pm \, c_B$ (C) and $q_3 = \pm \, c_B^2$ (D). The results of the solution behavior consideration in these cases are presented in Tab.\ref{Tab:SolPhys-z4a}. For generality both positive and negative $c_B$ values were supposed, but $c_B > 0$ lead to unstable solution violating NEC and should be excluded from the further
consideration. To continue possible scenarios narrowing the model thermodynamics investigation is needed.

\begin{table}[h!]
  \begin{center}
    \begin{tabular}{|c|c|c|c|c|c|}
      \hline
      Plot & $g(z)$ & $f_1(z)$ & $f_3(z)$ & $\phi(z)$ & $V(z)$ \\
      \hline
      \, 
      & $c_B < 0$ \ $c_B > 0$ & \, $c_B < 0$ \ $c_B >
      0$ \ & \ $c_B < 0$ \ $c_B > 0$ \ & \,$c_B < 0$ \ $c_B > 0$ \ &
      \, $c_B < 0$ \ $c_B > 0$ \ \\
      \hline
      A & stable unstable & stable \quad stable & NEC \ \ noNEC &
      stable unstable & stable unstable \\
      B & stable unstable & stable unstable & NEC \ \ noNEC & stable
      unstable & stable unstable \\
      C & stable unstable & stable unstable & NEC \ \ noNEC & stable
      unstable & stable unstable \\
      D & stable unstable & stable \quad stable & NEC \ \ noNEC &
      stable unstable & stable unstable \\
      \hline
    \end{tabular}
    \caption{Physicality of the system (\ref{eq:4.28}--\ref{eq:4.48})
      solution for the warp factor (\ref{eq:4.42}).}
    \label{Tab:SolPhys-z4a}
  \end{center}
\end{table}

\section{Thermodynamics and the 1-st order phase transition}\label{thermodynamics}

For the metric (\ref{eq:2.04}) and the warp factor (\ref{eq:4.42})
temperature and entropy are \cite{Arefeva:2023jjh}:

\begin{gather}
  \begin{split}
    T &= \cfrac{|g'|}{4 \pi} \, \Bigl|_{z=z_h} 
    = \left|
      - \, \cfrac{e^{(2R_{gg}-c_B)\frac{z_h^2}{2}+3(p-c_B
          \, q_3)z_h^4} \,
        z_h^{1+\frac{2}{\nu}}}{4 \pi \, \tilde I_1(z_h)} \, \times \right. \\
        &\times \left. \left[ 
        1 - \cfrac{\mu^2 \bigl(2 R_{gg} + c_B (q_3 - 1) \bigr) 
          \left(e^{(2 R_{gg} + c_B (q_3 - 1))\frac{z_h^2}{2}} \tilde
            I_1(z_h) - \tilde I_2(z_h) \right)}{L^2 \left(1 -
            e^{(2R_{gg}+c_B(q_3-1))\frac{z_h^2}{2}}
          \right)^2} \right] \right|,
  \end{split}\label{eq:4.49} \\
  \begin{split}
    \Tilde{I}_1(z) &= \int_0^z
    e^{\left(2R_{gg}-3c_B\right)\frac{\xi^2}{2}+3 (p-c_B \, q_3) \xi^4}
    \xi^{1+\frac{2}{\nu}} \, d \xi, \\
    \Tilde{I}_2(z) &= \int_0^z
    e^{\bigl(2R_{gg}+c_B\left(\frac{q_3}{2}-2\right)\bigr)\xi^2+3 (p-c_B
    \, q_3) \xi^4} \xi^{1+\frac{2}{\nu}} \, d \xi,
  \end{split}\label{eq:4.44} \\
  s = \cfrac{1}{4} \left( \cfrac{L}{z_h} \right)^{1+\frac{2}{\nu}}
  e^{-(2R_{gg}-c_B)\frac{z_h^2}{2}-3(p-c_B \, q_3)z_h^4}. \label{eq:4.50}
\end{gather}

\begin{figure}[t!]
  \centering
  \includegraphics[scale=1]{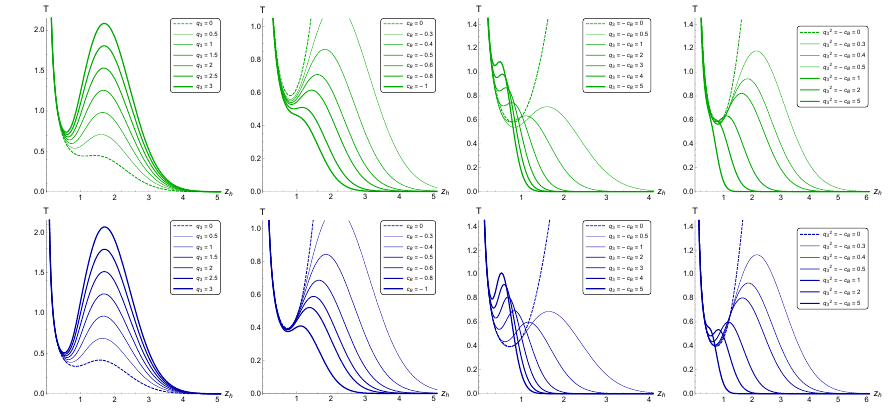} \\
  \ A \hspace{90pt} B \hspace{90pt} C \hspace{90pt} D \
  \caption{Temperature $T(z_h)$ in magnetic field with different $q_3$
    for $c_B = - \, 0.5$ (A), with different $c_B$ for $q_3 = 0.5$
    (B), for different $q_3 = - \, c_B$ (C), for different $q_3^2 = -
    \, c_B$ (D) for $\nu = 1$ (1-st line)
    and $\nu = 4.5$ (2-nd line), $\mu = 0$,
    $R_{gg} = 1.16$, $p = 0.273$.}
  \label{Fig:Tzh-q3cB-mu0-z4a}
  \ \\
  \centering
  \includegraphics[scale=1]{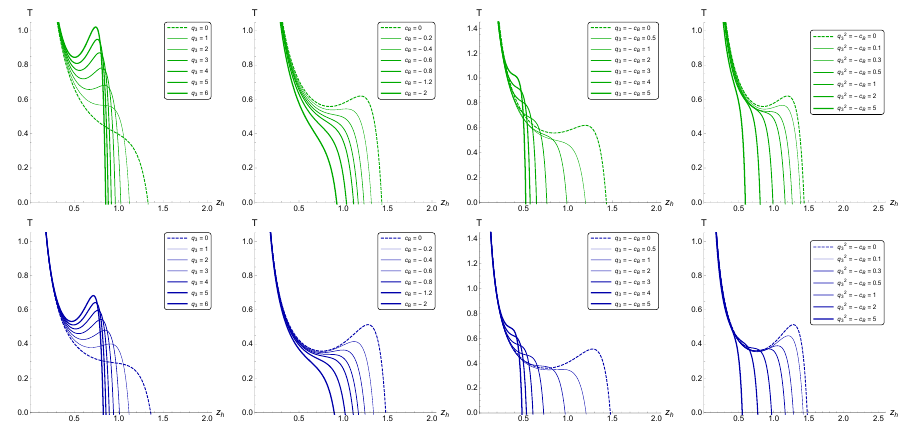} \\
  \ A \hspace{90pt} B \hspace{90pt} C \hspace{90pt} D \
  \caption{Temperature $T(z_h)$ in magnetic field with different $q_3$
    for $c_B = - \, 0.5$ (A), with different $c_B$ for $q_3 = 0.5$
    (B), for different $q_3 = - \, c_B$ (C), for different $q_3^2 = -
    \, c_B$ (D) for $\nu = 1$ (1-st line)
    and $\nu = 4.5$ (2-nd line), $\mu = 0.5$,
    $R_{gg} = 1.16$, $p = 0.273$.}
  \label{Fig:Tzh-q3cB-mu05-z4a}
\end{figure}
\begin{figure}[t!]
  \centering
  \ \\
  \includegraphics[scale=1]{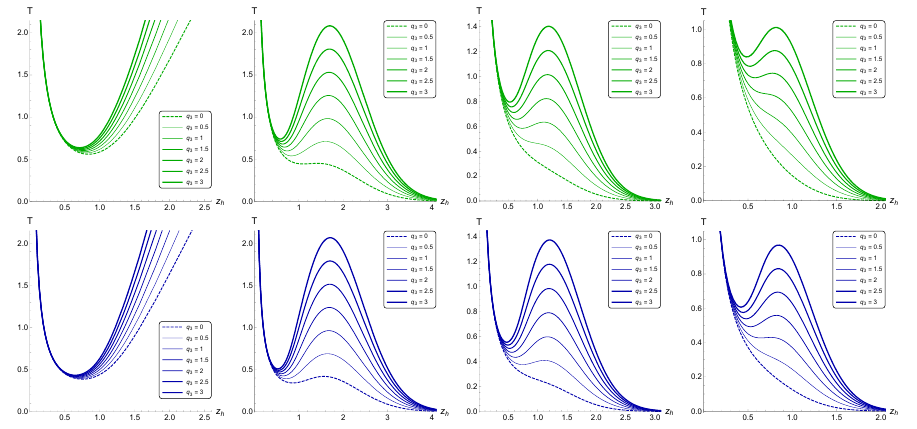} \\
  \ A \hspace{90pt} B \hspace{90pt} C \hspace{90pt} D \
  \caption{Temperature $T(z_h)$ in magnetic field with different $q_3$
    for $c_B = - \, 0.1$ (A), $c_B = - \, 0.5$ (B), $c_B = - \, 1$
    (C), $c_B = - \, 2$ (D) for $\nu = 1$ (1-st line)
    and $\nu = 4.5$ (2-nd line),
    $\mu = 0$, $R_{gg} = 1.16$, $p = 0.273$.}
  \label{Fig:Tzhq3-cB-mu0-z4a}
  \ \\
  \includegraphics[scale=1]{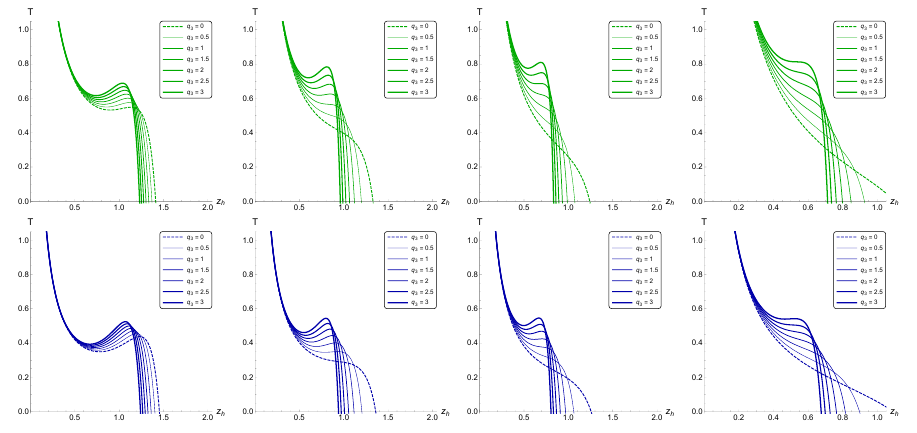} \\
  \ A \hspace{90pt} B \hspace{90pt} C \hspace{90pt} D \
  \caption{Temperature $T(z_h)$ in magnetic field with different $q_3$
    for $c_B = - \, 0.1$ (A), $c_B = - \, 0.5$ (B), $c_B = - \, 1$
    (C), $c_B = - \, 2$ (D) for $\nu = 1$ (1-st line)
    and $\nu = 4.5$ (2-nd line),
    $\mu = 0.5$, $R_{gg} = 1.16$, $p = 0.273$.}
  \label{Fig:Tzhq3-cB-mu05-z4a}
\end{figure}

\newpage

Fig.\ref{Fig:Tzh-q3cB-mu0-z4a}--\ref{Fig:Tzh-q3cB-mu05-z4a} show, that
direct magnetic catalysis within the current model can be expected
for the fixed $c_B$ (A), for $q_3 = - \, c_B$ (C) and in rather limited way for $q_3^2 = - \, c_B$ (D). We prefer to investigate fixed $c_B$ particular case, but others may be worth considering as well.

On Fig.\ref{Fig:Tzhq3-cB-mu0-z4a} $T(z_h,q_3)$ for different fixed magnetic parameter values $c_B = - \, 0.1$ (A),  $c_B = - \, 0.5$ (B),  $c_B = - \, 1$ (C) and  $c_B = - \, 5$ (D). The larger absolute $c_B$ value is, the lesser gap between temperature local maximum and minimum is. Besides, for $c_B \le - \, 1$ (C,D) zero and small magnetic field $q_3$ don't seem to allow 1-st order phase transition, that is also suppressed by larger chemical potential (
Fig.\ref{Fig:Tzhq3-cB-mu05-z4a}). 

\begin{figure}[t!]
  \centering
  \includegraphics[scale=1]{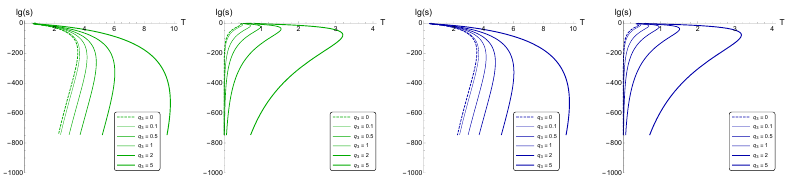} \\
  \ A \hspace{80pt} B \hspace{80pt} C \hspace{80pt} D \
  \caption{Entropy $s(T,q_3)$ in magnetic field for isotropic case $\nu = 1$ for $c_B = - \, 0.1$ (A), $c_B = - \, 0.5$ (B) and anisotropic case
    $\nu = 4.5$ for $c_B = - \, 0.1$ (C) and $c_B = - \, 0.5$ (D); $\mu = 0$, $R_{gg} = 1.16$, $p = 0.273$.}
  \label{Fig:sTq3-mu0-z4a}
\end{figure}

\begin{figure}[t!]
 \centering 
 \includegraphics[scale=1]{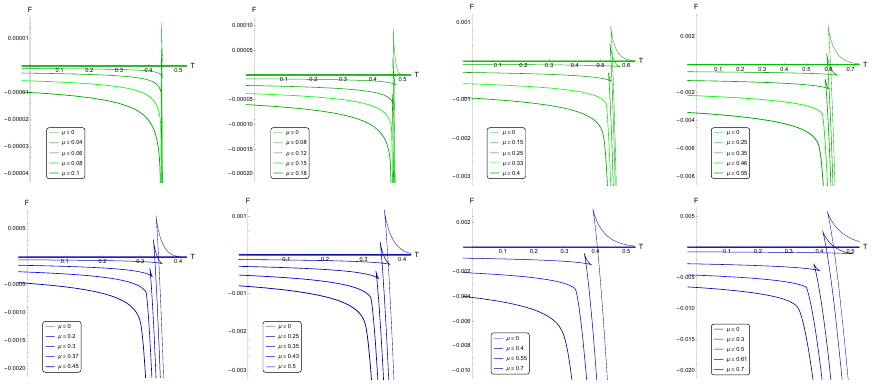} \\
 A \hspace{80pt} B \hspace{80pt} C \hspace{80pt} D
 \caption{Free energy $F(T,\mu)$ in magnetic field $q_3 = 0$ (A),
   $q_3 = 0.1$ (B), $q_3 = 0.5$ (C) and $q_3 = 1$ (D) 
   for isotropic case $\nu = 1$ (1-st line) and anisotropic case $\nu = 4.5$ (2-nd line), $c_B = - \, 0.5$, $R_{gg} = 1.16$, $p = 0.273$.}
 \label{Fig:FTmu-q3-nu1-p0273-cB-05-z4a}
\end{figure}

\begin{figure}[t!]
 \centering
 \includegraphics[scale=1]{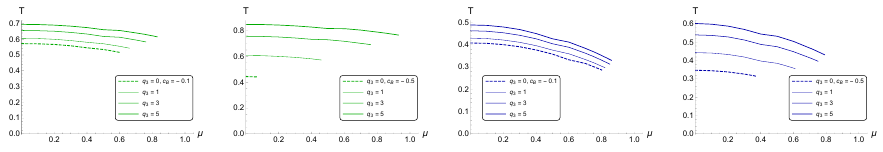} \\
 A \hspace{90pt} B \hspace{90pt} C \hspace{90pt} D \\
 \caption{Phase diagram for different magnetic field $q_3$ in asitropic case $\nu = 1$ (A,B) and anisotropic case $\nu = 4.5$ (C,D), $c_B = - \, 0.1$ (A,C) and $c_B = - \, 0.5$ (B,D); $R_{gg} = 1.16$.}
 \label{Fig:Tmuq3-cB-z4a}
\end{figure}

Entropy demonstrates the multi-valued behavior as the temperature does (Fig.\ref{Fig:sTq3-mu0-z4a}), thus making the collapse from small BH to large, corresponding to confinement-decon\-fine\-ment phase transition according to holographic dictionary, possible. The free energy
\begin{gather}
  F = - \int s \, d T = \int_{z_h}^{\infty} s \, T' dz \label{eq:5.36}
\end{gather}
expectedly displays the self-intersection form, the so-called ``swallow-tail'' (Fig.\ref{Fig:FTmu-q3-nu1-p0273-cB-05-z4a}), whose position determines the 1-st order phase transition (Fig.\ref{Fig:Tmuq3-cB-z4a}). The direct magnetic catalysis is clearly seen. Larger $c_B$ absolute value leads to stronger phase transition temperature dependence on $q_3$ and can even suppress the 1-st order phase transition for small magnetic field. On the other hand anisotropy stabilizes the 1-st order phase transition, but lowers its temperature.


\newpage

\section{Temporal Wilson loops and phase diagram}\label{twl}

\begin{figure}[t!]
  \centering
  \includegraphics[scale=1]{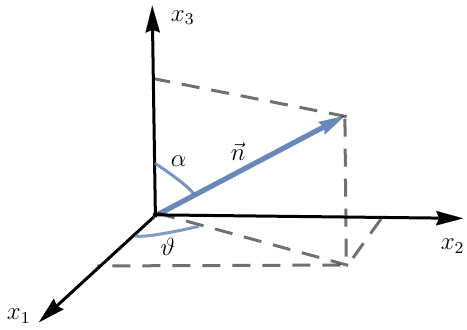}
  \caption{World sheet parametrization for the Wilson loop orientation
  vector $\vec n$.}
  \label{WLparam}
\end{figure}

To calculate the expectation value of the temporal Wilson loop
\begin{gather}
  W[C_\vartheta] = e^{-S_{\vartheta,t}}, \label{eq:5.37}
\end{gather}
oriented along vector $\vec n$
\begin{gather}
  n_{x1} = \cos \vartheta \sin \alpha, \quad
  n_{x2} = \sin \vartheta \sin \alpha, \quad 
  n_{x3} = \cos \alpha, \label{eq:5.38}
\end{gather}
we use our metric (\ref{eq:2.04}) as a background:
\begin{gather}
  ds^2 = G_{\mu\nu}dx^{\mu}dx^{\nu}
  = \cfrac{L^2}{z^2} \ \fb_s(z) \left[
    - \, g(z) dt^2 + \fg_1 dx^2 + \fg_2 dx_2^2 + \fg_3 dx_3^2
    + \cfrac{dz^2}{g(z)} \right], \label{eq:5.39}
\end{gather}
where $\fb_s(z) = \fb(z) \exp\bigl(\sqrt{2/3} \ \phi(z) \bigr)$ is the 
model warp-factor in string frame and $\fg_i$ are the corresponding
$G_{\mu\nu}$-metric components \cite{ARS-2019}. Following the holographic approach we
calculate the value of the Nambu-Goto action for test string in our
background in string frame:
\begin{gather}
  S = \frac{1}{2 \pi \alpha'} \int d\xi^{0} \, d\xi^{1} \sqrt{- \det
    h_{\alpha\beta}}, \qquad
  h_{\alpha\beta} = G_{\mu\nu} \, \partial_{\alpha}
  X^{\mu} \, \partial_{\beta} X^{\nu}. \label{eq:5.40}
\end{gather}
The world sheet is parameterised as (Fig.\ref{WLparam})
\begin{gather}
  \begin{split}
    X^{0} \equiv t = \xi^0, \quad
    X^{1} &\equiv x = \xi^1 \cos\vartheta \sin \alpha, \quad
    X^{2} \equiv y_{1} = \xi^1\sin\vartheta \sin \alpha, \\
    X^{3} &\equiv y_{2} = \xi^1 \cos \alpha, \qquad \ \ \,
    X^{4} \equiv z = z(\xi^1).
  \end{split}
    \label{eq:5.41}
\end{gather}

Let us denote $\xi \equiv \xi^1$, rewrite Nambu-Goto action
(\ref{eq:5.40})
\begin{gather}
  \begin{split}
    S &= - \, \cfrac{\tau}{2 \pi \alpha'} \ {\cS}, \quad \ \
    {\cal S} = \int  _{-\ell/2}^{\ell/2}\,d\xi \,
    M\left(z(\xi)\right) \, \sqrt{{\cal F}(z\left(\xi)\right) +
      \left(z'(\xi)\right)^2}, \\
    M\left(z(\xi)\right) &= \cfrac{\fb_s\left(z(\xi)\right)}{z^2(\xi)}
    \, ,  \qquad \tau = \int d \xi^0, \\
    {\cal F}\left(z(\xi)\right) &= g\left(z(\xi^1)\right) \left(\fg_1
      \cos^2\vartheta \sin^2 \alpha + \fg_2 \sin^2\vartheta \sin^2
      \alpha + \fg_3 \cos^2 \alpha \right)
  \end{split} \label{eq:5.42}
\end{gather}
and introduce the effective ``potential''
\begin{gather}
  \begin{split}
    {\cal V}\left(z(\xi)\right) 
    &\equiv M\left(z(\xi)\right) \sqrt{{\cal F}\left(z(\xi)\right)} =
    \\
    &= \cfrac{\fb_s\left(z(\xi)\right)}{(z^2(\xi))} \,
    \sqrt{g\left(z(\xi)\right) \left(\fg_1 \cos^2\vartheta \sin^2
        \alpha + \fg_2 \sin^2\vartheta \sin^2 \alpha + \fg_3 \cos^2
        \alpha \right)}.
  \end{split}\label{eq:5.43}
\end{gather}

The action ${\cal S}$ \eqref{eq:5.42} defines the dynamical system
with a dynamic variable $z = z(\xi)$ and time~$\xi$ (see for example
\cite{ARS-2019}). This system has the first integral
\begin{gather}
  \frac{M(z(\xi)){\cal F}(z(\xi))}{\sqrt{{\cal
      F}(z(\xi))+(z'(\xi))^2}} = {\cal I}. \label{eq:5.44}
\end{gather}
From (\ref{eq:5.44}) we find the ``top'', or the turning point,
${z_{0}}$ (the closed position of the minimal surface to the horizon),
where ${z'(\xi) = 0}$:
\begin{gather}
  M(z_{0}) \sqrt{F(z_{0})} = {\cal I}. \label{eq:5.45}
\end{gather}
Representations for the string length $\ell$ and the action ${\cal S}$ \eqref{eq:5.42} are given by $z^{\prime}$ from (\ref{eq:5.44}):
\begin{gather}
  \frac\ell2 =
  \int_0^{z_0} \frac{1}{\sqrt{{\cal F}(z)}} \
  \frac{dz}{\sqrt{\frac{{\cal V}^2(z)}{{\cal V}^2(z_0)} -1
    }}, \quad 
  \frac{{\cal S}}{2} =
  \int_\epsilon^{z_0}\frac{M(z) \, dz}{\sqrt{1-\frac{{\cal V}^2(z_0)}{{\cal
          V}^2(z)}}}. \label{eq:5.47}
\end{gather}
To obtain the string tension we have to calculate asymptotic of $\cal
S$ for $\ell\to \infty$. If the stationary point of ${\cal V}(z)$
exists in the region $0 < z < z_h$, 
\begin{gather}
  {\cal V}^\prime\Big|_{z_{DW}} = 0, \label{eq:5.48}
\end{gather}
we call this point  a dynamical wall (DW) point and  take the top
point $z_0$ equal to the DW position, $z_0 = z_{DW}$. Near this point
we get 
\begin{gather}
  \ell \underset{z\to z_{DW}}{\sim} \frac{1}{\sqrt{F(z_{DW})}} \,
  \sqrt{\frac{{\cal V}(z_{DW})}{{\cal V}''(z_{DW})}} \, \log
  (z - z_{DW}) \underset{z \to z_{DW}} \to \infty, \label{eq:5.49} \\
  {\cS} \underset{z \to z_{DW}}{\sim} M(z_{DW}) \, \sqrt{\frac{{\cal
        V}(z_{DW})}{{\cal V}''(z_{DW})}} \, \log (z -
  z_{DW}). \label{eq:5.50}
\end{gather}
Hence  for the  string tension we get
\begin{gather}
  {\cS} \sim \sigma_{DW} \cdot \ell\,,\qquad 
  \sigma_{DW} = M(z_{DW}) \, \sqrt{F(z_{DW})}. \label{eq:5.51}
\end{gather}
If there is no stationary point of ${\cal V}(z)$ in the region $0 < z
< z_h$, the string tension becomes 
\begin{gather}
  \sigma_h = M(z_h) \, \sqrt{F(z_h)}. \label{eq:5.52}
\end{gather}

\begin{figure}[t!]
  \centering 
  \includegraphics[scale=1]{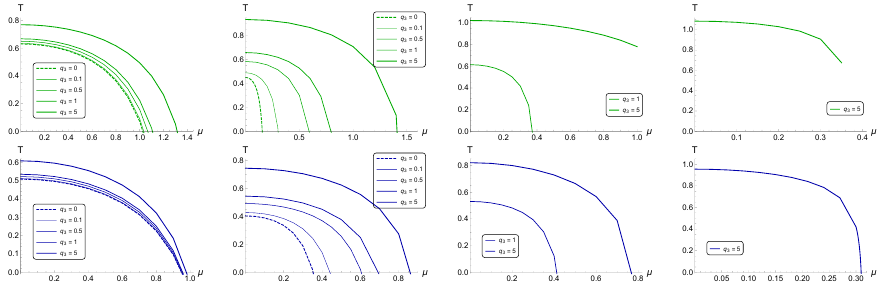} \\
  A \hspace{90pt} B \hspace{90pt} C \hspace{90pt} D
  \caption{Wilson loop phase transition curves $T(\mu)$ for different
    magnetic field $q_3$ for $c_B = - \, 0.1$ (A), $c_B = - \, 0.5$ (B), $c_B = - \, 1$ (C) and $c_B = - \, 5$ (D) in primary isotropic case $\nu = 1$ (1-st line) and primary anisotropic case $\nu = 4.5$ (2-nd line), $R_{gg} = 1.16$, $p = 0.273$.} 
  \label{Fig:WLmuq3-cB-z4a} 
  \ \\
  \includegraphics[scale=1]{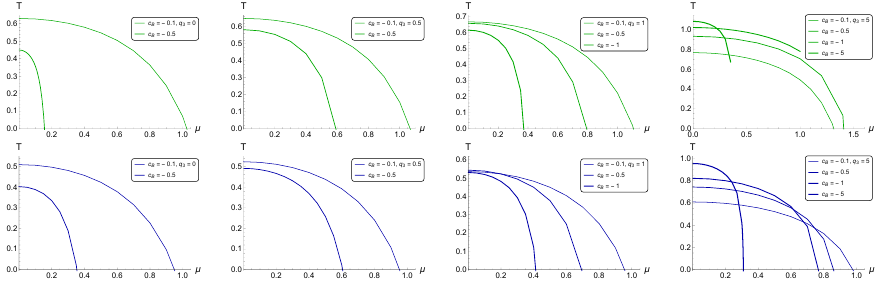} \\
  A \hspace{90pt} B \hspace{90pt} C \hspace{90pt} D 
  \caption{Wilson loop phase transition curves $T(\mu)$ for different
    magnetic field metric coefficient $c_B$ in magnetic field $q_3 = 0$ (A), $q_3 = 0.5$ (B), $q_3 = 1$ (C) and $q_3 = 5$ (D) in primary isotropic case $\nu = 1$ (a-st line) and primary anisotropic case $\nu = 4.5$ (w-nd line), $R_{gg} = 1.16$, $p = 0.273$.} 
  \label{Fig:WLmucB-q3-z4a}
\end{figure}
Thus  the turning points  for Wilson loops WL$x_1$,  WL$x_2$ and WL$x_3$, 
oriented along $x_1$, $x_2$ and $x_3$ axes, respectively, are defined 
by equations \cite{ARS-Heavy-2020, ARS-Light-2022}:

\begin{gather}
  \hspace{-90pt}
  \mbox{WL}x_1: \quad - \, \cfrac43 \, R_{gg} z 
        - 8 (p - c_B q_3) z^3 
        + \sqrt{\cfrac23} \ \phi'(z) + \cfrac{g'}{2 g} 
        - \cfrac{2}{z} \ \Big|_{z = z_{DWx_1}} \hspace{-15pt} = 0, \label{eq:5.53} \\
  \hspace{-37pt}
  \mbox{WL}x_2: \quad - \, \cfrac43 \, R_{gg} z 
        - 8 (p - c_B q_3) z^3
        + \sqrt{\cfrac23} \ \phi'(z) + \cfrac{g'}{2 g} 
        - \cfrac{\nu + 1}{\nu z} \ \Big|_{z = z_{DWx_2}}
        \hspace{-15pt} = 0, \label{eq:5.54} \\
  \hspace{-5pt}
  \mbox{WL}x_3: \quad - \, \cfrac43 \, R_{gg} z 
        - 8 (p - c_B q_3) z^3
        + \sqrt{\cfrac23} \ \phi'(z) + \cfrac{g'}{2 g} 
        - \cfrac{\nu + 1}{\nu z} + c_B z \ \Big|_{z = z_{DWx_3}}
        \hspace{-15pt} = 0. 
  \label{eq:5.55}
\end{gather}

\begin{figure}[b!]
  \centering 
  \includegraphics[scale=1]{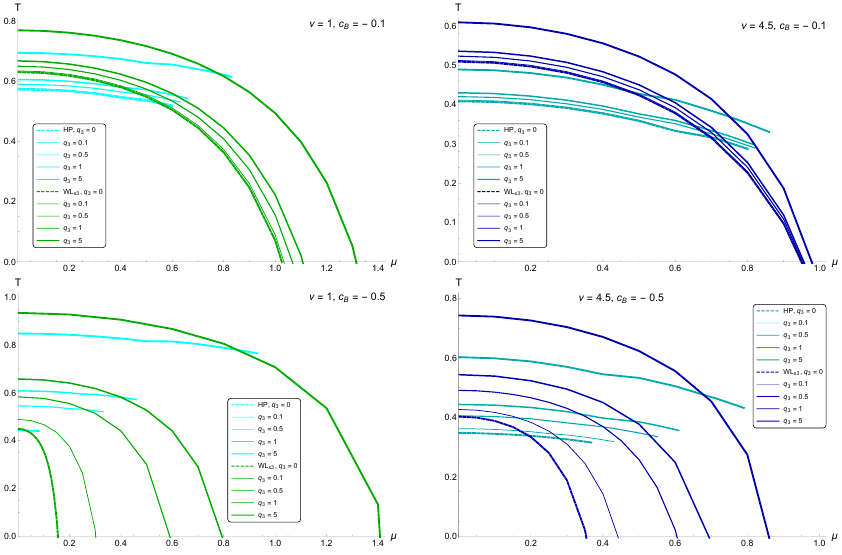} \\
  A \hspace{200pt} B 
  \caption{Phase diagram $T(\mu)$ in magnetic field for $c_B = - \, 0.1$ (1-st line) and $c_B = - \, 0.5$ (2-nd line) in primary isotropic case $\nu = 1$ (A) and anisoropic case $\nu = 4.5$ (B); $R_{gg} = 1.16$, $p = 0.273$.} 
  \label{Fig:PDmuq3-cB-z4a}
\end{figure}

It is obvious, that for $c_B = 0$ expressions in (\ref{eq:5.54}) and
(\ref{eq:5.55}) become equal, and putting $\nu = 1$ reduces all
equations (\ref{eq:5.53})-(\ref{eq:5.55}) to (\ref{eq:5.53}). The
lines in the plane $(\mu,z_h)$ and therefore in the plane $(\mu,T)$,
where equations (\ref{eq:5.53})-(\ref{eq:5.55}) stop having real
solutions, are interpreted as lines of Wilson phase
transition. Locations in the $(\mu,T)$-plane of the phase transition
lines for Wilson loops WL$x_1$ and WL$x_2$  were already considered
\cite{AR-2018, ARS-2019}.

Confinement-deconfinement phase transition, determined by temporal Wilson loops (a crossover), also undergoes the direct magnetic catalysis if the magnetic field $q_3$ increases (Fig.\ref{Fig:WLmuq3-cB-z4a}). For larger $c_B$ absolute values the effect is the opposite (Fig.\ref{Fig:WLmucB-q3-z4a}) and is strong field can look rather complicated (Fig.\ref{Fig:WLmucB-q3-z4a}.D). The crossover suppression for small magnetic field $q_3$ and large secondary anisotropy metric coefficient $c_B$ absolute value is confirmed.

The phase structure typical for the heavy quarks \cite{AR-2018, ARS-2019, ARS-Heavy-2020} is clearly seen (Fig.\ref{Fig:PDmuq3-cB-z4a}). Larger $c_B$ absolute value enforces the magnetic catalysis effect, prolongs the 1-st order phase transition contribution  to larger chemical potential values and notably shifts the confinement-deconfinement $\mu_{max}$: $T(\mu_{max}) = 0$. Anisotropy not only lowers the phase transition temperature, but also weakens the effects mentioned above, thus stabilising the phase diagram.

\begin{figure}[t!]
 \centering 
 \includegraphics[scale=1]{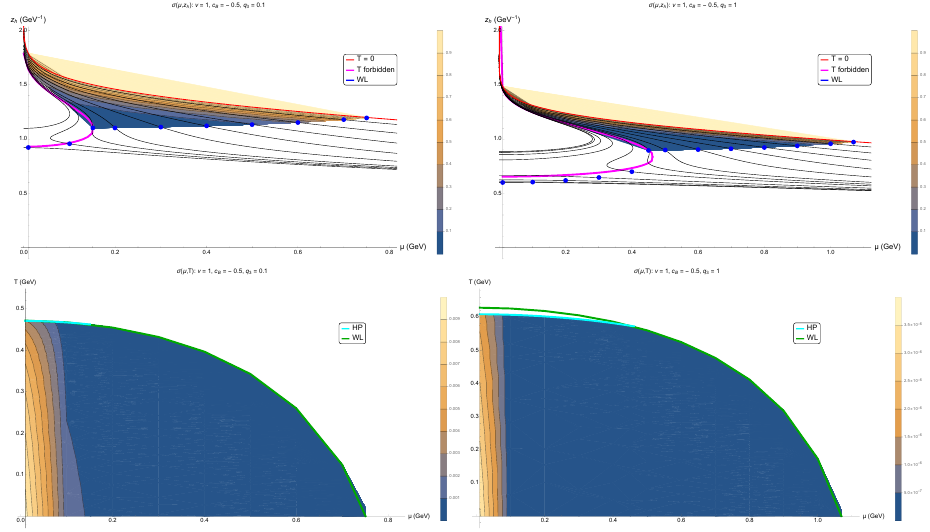} \\
 A \hspace{200pt} B 
 \caption{String tension $\sigma(\mu,z_h)/\sigma_0$ (1-st line) and  $\sigma(\mu,T)$ (2-nd line) in magnetic field $q_3 = 0.1$ (A) and $q_3 = 1$ (B);  $R_{gg} = 1.16$, $p = 0.273$, $\nu = 1$, $c_B = - \, 0.5$.}
 \label{Fig:sigma-q3-cB-05-z4a}
\end{figure}

The string tension consideration requires to determine the scalar field (dilaton) boundary $z_0$: $\phi_{z_0} = 0$. In previous works we used $z_0 = e^{-z_h/4} + 0.1$ for heavy quarks and $z_0 = 10 e^{-z_h/4} + 0.1$ for light quarks model \cite{Arefeva:2025xtz}. In current research the dilaton boundary was revised, and the expression considered optimal and used for the string tension calculations is $z_0 = 3 e^{-z_h/2} + 0.1$. The results for isotropic case are presented on Fig.\ref{Fig:sigma-q3-cB-05-z4a}. Black curves on $(\mu,z_h)$-plane (1-st line) are isotherms and the red curve corresponds to $T = 0$; magenta line shows the border of the forbidden area, to where no stable transition is possible; blue dots mark the position of the crossover. The string tension value is normalized to $\sigma_0 = \sigma(T = 0)$ and presented by the density plot. 2-nd line of Fig.\ref{Fig:sigma-q3-cB-05-z4a} shows the string tension behavior on the $(\mu,T)$-plane. One can see that $\sigma$ quickly decreases with chemical potential and, significantly slower, for temperature. The 1-st order phase transition and the crossover limit the area where the string tension exists, i.e. the area of confinement. In larger magnetic field string tension is weaker ($1$ order of $q_3$ leads to about $3$ orders difference in $\sigma$) and decreases to near-zero values faster.

\section{Discussion}\label{conclusions}

In this work temporal Wilson loops for the holographic heavy quarks model \cite{Arefeva:2023jjh} are investigated. The confinement-deconfinement (crossover) phase transition and the string tension in magnetic field are obtained. It is shown that there is a number of opportunities to parametrize magnetic field and the secondary anisotropy caused by it within the current model. Thus the presented holographic description is proven to be rather flexible for fitting experimental and Lattice data.

The string tension behavior in magnetic field and in the presence of primary anisotropy needs further detailed investigation. In particular the question on the dilaton boundary retains its importance in connection with the problem of a generalized description of heavy and light quarks for the creation of a hybrid model capable of describing quarks of different compositions. Let us note that the string tension itself is a step towards the Cornell potential consideration that should be performed for the current model as well as for the other heavy quarks version \cite{Rannu:2024vrq} and the light quarks model \cite{ARS-Light-2022, Arefeva:2022bhx}.

\section{Acknowledgments}

The author greatly thanks I. Ya. Aref'eva and P. S. Slepov for fruitful
discussions. This work was performed within the scientific project
\textnumero~FSSF-2023-0003 and the ``BASIS'' Science Foundation grant
\textnumero~22-1-3-18-1.

\appendix 


\section{Appendix. Solution}\label{appendixA}

\begin{gather}
  A_t (z) = \mu \left( 1 - \cfrac{1 - {e^{(2
          R_{gg}+c_B(q_3-1))\frac{z^2}{2}}}}{1 - {e^{(2
          R_{gg}+c_B(q_3-1))\frac{z_h^2}{2}}}}
  \right), \label{eq:4.28} \\
  g(z) = e^{c_B z^2} \Biggl[ 1 -
  \cfrac{\Tilde{I}_1(z)}{\Tilde{I}_1(z_h)}
  + \cfrac{\mu^2 \bigl(2 R_{gg} + c_B (q_3 - 1) \bigr)
    \Tilde{I}_2(z)}{L^2 \left(1 - e^{(2
        R_{gg}+c_B(q_3-1))\frac{z_h^2}{2}}
    \right)^2} \left( 1 - \cfrac{\Tilde{I}_1(z)}{\Tilde{I}_1(z_h)} \,
    \cfrac{\Tilde{I}_2(z_h)}{\Tilde{I}_2(z)} \right)
  \Biggr], \label{eq:4.43} \\
  \begin{split}
    \phi(z) = \int_{z_0}^z \cfrac{\sqrt{2}}{\nu \xi} \ 
    &\Biggl[
    2 (\nu - 1)
    + \bigl(6 R_{gg} \nu + ( 2 - 3 \nu ) c_B \bigr) \nu \xi^2
    + \left( \cfrac{4}{3} \, R_{gg}^2 - c_B^2 + 60 (p-c_B
      \, q_3) \right) \nu^2 \xi^4 \, + \\
    &\qquad \qquad \, + 16 R_{gg} (p-c_B \, q_3) \nu^2 \xi^6
    + 48 (p-c_B \, q_3)^2 \nu^2 \xi^8 \Biggr]^{1/2} d \xi,
  \end{split}\label{eq:4.45} \\
  \begin{split}
    f_1(\phi(z)) &= 2 \left( \cfrac{z}{L} \right)^{2-\frac{4}{\nu}} \!\!\!
    e^{-2(R_{gg}-3c_B)z^2/3-2(p-c_B \, q_3)z^4} \, \cfrac{\nu -
      1}{\nu^2 q_1^2 z^2} \, \times \\
    &\times \Biggl\{
    \left( 2 (1 + \nu)
      + (2 R_{gg} - 3 c_B) \nu z^2
      + 12 (p-c_B \, q_3) \nu z^4 \right) \times \\
    &\times \left[ 1 - \cfrac{\tilde I_1(z)}{\tilde I_1(z_h)}
      + \cfrac{\mu^2 \bigl( 2 R_{gg} + c_B (q_3 - 1) \bigr)}{L^2 \left(
          1 - e^{\left(2R_{gg} + c_B (q_3 - 1)\right) z_h^2/2} \right)^2}
      \left( \tilde I_2(z_h) - \cfrac{\tilde I_1(z_h)}{\tilde
          I_1(z_h)} \ \tilde I_2(z) \right) \right] + \\
    &+ \cfrac{e^{(2R_{gg}-3c_B)z^2/2+3(p-c_B \, q_3)z^4} \nu 
      z^{2+\frac{2}{\nu}}}{\tilde I_1(z_h)}
    \left[ 1 - \cfrac{\mu^2 \bigl(2 R_{gg} + c_B (q_3 - 1) \bigr)}{L^2
        \left( 1 - e^{\left(2R_{gg} + c_B (q_3 - 1) \right) z_h^2/2}
        \right)^2} \right. \times \\
      &\times \left. \left( e^{\left(2 R_{gg} + c_B (q_3 - 1) \right) z^2/2} \tilde
        I_1(z_h) - \tilde I_2 
        (z_h) \right) \right] \Biggr\},
  \end{split} \label{eq:4.46} 
\end{gather}
\begin{gather}
  \begin{split}
    f_3(\phi(z)) = &- 2 \left( \cfrac{z}{L} \right)^{-\frac{2}{\nu}}c \!\!\!
    e^{-(2R_{gg}-3c_B)z^2/3-2(p-c_B \, q_3)z^4} \, \cfrac{c_B}{\nu
      q_3^2} \, \times \\
    &\times \Biggl\{ \left( 2 + (2 R_{gg} - 3 c_B) \nu z^2 + 12 (p-c_B \, q_3)
      \nu z^4 \right) \times \\
    &\times \Biggl[ 1 - \cfrac{\tilde I_1(z)}{\tilde I_1(z_h)}
    + \cfrac{\mu^2 \bigl(2 R_{gg} + c_B (q_3 - 1)\bigr)}{L^2 \left(
        1 - e^{(2R_{gg}+c_B(q_3-1))z_h^2/2} \right)^2}
    \ \left( \tilde I_2(z) - \cfrac{\tilde I_1(z)}{\tilde I_1(z_h)} \
      \tilde I_2(z_h) \right) \Biggr] + \\
    &+ \cfrac{e^{(2R_{gg}-3c_B)z^2/2+3(p-c_B \, q_3)z^4} \nu
      z^{2+\frac{2}{\nu}}}{\tilde I_1(z_h)}
    \left[ 1 - \cfrac{\mu^2 \bigl(2 R_{gg} + c_B (q_3 - 1)\bigr)}{L^2
        \left(1 - e^{(2R_{gg}+c_B(q_3-1))z_h^2/2} \right)^2} \right. \times \\
      &\times \left. \left( e^{(2R_{gg}+c_B(q_3-1))z^2/2} \tilde I_1(z_h) - \tilde I_2
        (z_h) \right) \right] \Biggr\},
  \end{split} \label{eq:4.47} \\
  \begin{split}
    V(\phi(z)) = &- \cfrac{e^{(2R_{gg}+3c_B)z^2/3+2(p-c_B \,
        q_3)z^4}}{L^2 \nu^2} \,
    \Biggl\{ \left( 2 \, (1 + 3 \nu + 2 \nu^2) + \bigl( 2 R_{gg} (3 +
      2 \nu) - (7 + 6 \nu) c_B \bigr) \nu z^2 \right. + \\
    &+ 2 \left( (2 R_{gg}^2 - 5 R_{gg} c_B + 3 c_B^2) \nu + 18 (p-c_B
      \, q_3) \right) \nu z^4 + \\
    &+ 12 \left. (4 R_{gg} - 5 c_B) (p-c_B \, q_3) \nu^2 z^6 + 144 (p-c_B \, q_3)^2 \nu^2 z^8 \right) \times \\
    &\times \Biggl[ 1 - \cfrac{\tilde I_1(z)}{\tilde I_1(z_h)}
    + \cfrac{\mu^2 \bigl(2 R_{gg} + c_B (q_3 - 1)\bigr)}{L^2 \left(
        1 - e^{(2R_{gg}+c_B(q_3-1))z_h^2/2} \right)^2}
    \ \left( \tilde I_2(z) - \cfrac{\tilde I_1(z)}{\tilde I_1(z_h)} \
      \tilde I_2(z_h) \right) \Biggr] \ + \\
    &+ \left( 1 + 2 \nu + 2 \, (R_{gg} - c_B) \nu z^2 
      + 12 (p-c_B \, q_3) \nu z^4 \right)
    \cfrac{e^{(2R_{gg}-3c_B)z^2/2+3(p-c_B \, q_3)z^4} \nu
      z^{2+\frac{2}{\nu}}}{\tilde I_1(z_h)} \ \times \\
    &\times \left[ 1 - \cfrac{\mu^2 \bigl(2 R_{gg} + c_B (q_3 -
        1)\bigr)}{L^2 \left( 1 - e^{(2R_{gg}+c_B(q_3-1))z_h^2/2}
        \right)^2} \right. \times \\
    &\times \Biggl( \left( 1 - \cfrac{\left( 2 R_{gg} + c_B (q_3
              - 1) \right) \nu z^2}{2 \left( 1 + 2 \nu + 2 \, (R_{gg}
              - c_B) \nu z^2 + 12 (p-c_B \, q_3) \nu z^4 \right)}
        \right) \times \\
    &\times \left. e^{(2R_{gg}+c_B(q_3-1))z^2/2} \tilde I_1(z_h) - \tilde
        I_2(z_h) \Biggr) \right] \Biggr\}.
  \end{split} \label{eq:4.48}
\end{gather}

\end{document}